\documentclass[11pt]{article}
\usepackage{amsmath,amssymb,amsfonts}
\usepackage[english]{babel}
\usepackage{float}
\makeatletter
\@addtoreset{equation}{section}
\makeatother

\def\nn{\nonumber} \def\bd{\begin{document}} \def\ed{\end{document}}
\def\ds{\documentstyle}
\let\bm=\bibitem
\newcommand{\be}{\begin{equation}}
\newcommand{\ee}{\end{equation}}
\newcommand{\bea}{\setlength\arraycolsep{2pt} \begin{eqnarray}}
\newcommand{\eea}{\end{eqnarray}}
\newcommand{\hoch}[1]{$\, ^{#1}$}
\def\p{\partial}

\title{\large {\bf Generalized Komar currents for vector fields}}
\date{}

\author{Jun-Jin Peng$^{1,2}$\footnote{Corresponding author: pengjjph@163.com},
\quad Chang-Li Zou$^{1}\footnote{zoucl2019@163.com}$ \\ \\
\small \sl $^1$School of Physics and Electronic Science, Guizhou Normal University,\\
\small Guiyang, Guizhou 550001, People's Republic of China; \\
\small \sl  $^2$Guizhou Provincial Key Laboratory of Radio Astronomy and
Data Processing, \\
\small \sl Guizhou Normal University, \\
\small Guiyang, Guizhou 550001, People's Republic of China
}

\begin{document}

\maketitle
\vspace{20pt}

\begin{center}
\textbf{Abstract}
\end{center}

In this paper, on basis of three quadratic differential operators leaving
the form degree of an arbitrary differential form unchanged, that is, the
d'Alembertian operator and two combined ones from the Hodge coderivative
and the exterior derivative, the usual Komar current for a Killing vector
is formulated into another equivalent form. Then it is extended to
more general currents in the absence of the linearity in the Killing
vector field. Moreover, motivated by this equivalent of the usual Komar
current, we put forward a conserved current corresponding to a generic
vector with some constraint. Such a current
can be generalized to the one with higher-order derivatives of the vector.
The applications to some specific vector fields, such as the almost-Killing
vectors, the conformal Killing vectors and the divergence-free vectors,
are investigated. It is demonstrated that the above generalizations
of the Killing vector can be uniformly described by a second-order
derivative equation.

\voffset=-.90pt
\vspace{40pt}

\section{Introduction}\label{one}

General relativity, together with its various modifications, possesses a diffeomorphic
invariance in accordance with the requirement of the covariance principle for the
theory itself. Ordinarily, such a symmetry is generated by vector fields, particularly
by the Killing vectors. As a consequence, inspired with the famous Noether theorem,
one can naturally anticipate that each vector field has the potential to bring into a
conserved current, and further to give rise to a corresponding conserved charge.

In fact, a well-known significant example to demonstrate the above statement
is the appearance of the so-called Komar conserved current for general relativistic
spacetime manifolds \cite{Komar}, although it was found without the guidance of
the standard Noether approach.
By virtue of such a simple but useful current, the Komar integral was put forward.
Till now, it has occupied an important position in the definition of the conserved charges in
asymptotically flat spacetime. Nevertheless, when the Komar integral
is carried out to calculate the conserved charges of spacetimes that do not behave as
asymptotical flatness, it usually gives rise to the divergence problems. Not only that, it
sometimes fails to reproduce the anticipated values for all the conserved charges
even though they are finite \cite{Katfac2,PeKf2,JuSi}. Consequently, the usual
Komar current, as well as the Komar integral, deserves to be modified.
Actually, there has existed a lot of literature involved in its modifications and
generalizations. For example, see the works
\cite{PetFtc,JuFeng,LBBKofl,CleGal,KasKinHDG,BaZygrg,BNSnut,MielAgK,ACOTZ,BFFV,GiaSa,MaKiAdS,TamWin}.

According to the ordinary Komar current, one can observe that it merely consists of a
second-order derivative term of the vector field. Apart from this, the conventional Komar
current, as well as its various generalizations, is linear in the vector. In order to
provide interesting insights into the Komar current corresponding to the Killing vector,
we are going to take into account the non-linear generalization to the usual
Komar current under the assumption that the current depends at most upon the
second-order derivatives of the Killing vector and its linearity in the
vector is lost. In terms of such a generalized current, it may be possible to
build a meaningful formula for the conserved charges of gravity theories.

More generally, we may even go further, taking into account the construction
of the conserved currents associated to a generic vector field (specifically,
the conserved currents are restricted to depend only on the vector and the
metric tensor, as well as their derivatives). As a matter of fact, each
conserved current can be seen as a divergenceless vector. This gives us a
clue that the conserved currents, as well as the vectors involved in them, are
able to be understood as 1-forms, which can be manipulated by all the
operators within the framework of differential forms. Consequently, it is
feasible to obtain various currents by letting the Hodge star operation
$\star$ and the exterior derivative $d$, together with their combinations,
act on the vector fields. However, in order to ensure that the form degree
of the currents remains one, it is of great necessity to choose the operators that
can leave the form degree of the 1-form vector fields unaltered. Really, as what has been illustrated in the
work \cite{Ppformin}, for an arbitrary $p-$form $\textbf{F}$ in an
$n$-dimensional spacetime manifold with a metric $g_{\mu\nu}$ having a
Lorentzian signature $(-,+,+,\cdot\cdot\cdot)$, under the action of the
generally-covariant d'Alembertian operator $\Box$, as well
as the second-order operation $\hat{\delta}d$ or $d\hat{\delta}$
(the concrete forms for their action on $\textbf{F}$
can be found in the appendix \ref{appendA}), the form degree of the $p-$form
remains invariant. Here the codifferential (or the divergence operator) $\hat{\delta}$
is defined through $\hat{\delta}=(-1)^{np+n+1}\star d \star$, and its
action on $\textbf{F}$ leads to
$(\hat{\delta}\textbf{F})_{\mu_2\cdot\cdot\cdot\mu_p}
=\nabla^{\mu_1} F_{\mu_1\cdot\cdot\cdot\mu_p}$ (for details on the operators
$\hat{\delta}$, $\hat{\delta}d$ and $d\hat{\delta}$, see \cite{Ppformin} and
references therein). Furthermore, it will be proved below that all the three
quadratic differential operators $\Box$, $d\hat{\delta}$ and $\hat{\delta}d$
are the primary ones with the lowest differential order that can preserve
the form degree of any
differential form. According to this, they may be the ideal candidates to
generate conserved currents out of various vector fields. Therefore,
we arrive at the question whether the three operators
$\Box$, $d\hat{\delta}$ and $\hat{\delta}d$ can bring about some interesting
conserved currents by means of their action on a Killing vector field or
a generic one.

For the purpose of answering the above questions, under the condition that the
vector fields (including the Killing vectors) together with the conserved currents
are treated as 1-forms and the three objects $\Box$, $d\hat{\delta}$ and
$\hat{\delta}d$ are chosen as
the fundamental operators acting on them, we attempt to supply another way
to find out the possible conserved currents associated
with these vectors. The outcome will show that the usual Komar current can
be generalized to the ones containing non-linear terms of the Killing vectors.
Besides, the three operations together with their linear combination render
us to conveniently put forward the conserved currents made up of quadratic or
higher-order derivatives of any general vector field. Remarkably, by virtue
of the operator constructed from the linear combination of the three
elementary ones, the ordinary Komar current for the Killing vector is able
to be written as a novel equivalent form, while the equations encompassing
the second-order derivatives of the vector fields can be expressed as a
unified formulation.

The remainder of the present paper goes as follows. In section \ref{two},
we are going to investigate various possible conserved currents associated
with the Killing vector fields. Subsequently, the usual Komar current will
be generalized to the one with respect to an arbitrary vector by means of
the three differential operations $\Box$, $\hat{\delta}d$ and $d\hat{\delta}$.
In order to demonstrate the meanings of the current, a couple of examples
for some specific vector fields will be presented. In section \ref{three},
we shall pay attention to the extended currents that are dependent at most
of second-order derivatives of a general vector field, as well as the
currents consisting of higher-order derivative terms of the vector. The
last section is our discussions and remarks. The main results of this paper
will be summarized in appendix \ref{appendD}.

\section{Generalized Komar currents associated to vector fields}\label{two}

In the present section, with the help of all the three second-order differential
operations $\Box$, $\hat{\delta}d$ and $d\hat{\delta}$ (the introduction
for them is given in the appendix \ref{appendA}), it is of great convenience
for us to put the ordinary Komar current for a Killing vector into another novel
form. According to this, we shall put forward a more general conserved current
that is made up of terms proportional to at most second-order derivatives of the
Killing vector, as well as a conserved current with respect to arbitrary vector
fields. As some examples, the currents associated
with a couple of specific interesting vectors, such as the almost-Killing vectors, the
conformal Killing vectors and the divergence-free vector fields, are deduced. It
is worth noticing that it is postulated that the currents are merely dependent
of the vector fields and metric tensors, as well as their derivatives,
throughout this work.

\subsection{Komar currents in terms of the operations
$\Box$, $\hat{\delta}d$ and $d\hat{\delta}$}\label{ss21}

As a beginning, for simplicity, we take into account the ordinary Komar
current $\textbf{J}_K$
corresponding to a Killing vector $\xi^{\mu}$, which describes the infinitesimal
isometries of an $n$-dimensional spacetime manifold equipped with a
pseudo-Riemannian metric $g_{\mu\nu}$ and is determined by the well-known
Killing equation $2\nabla_{(\mu}\xi_{\nu)}=0$. The conserved Komar current
is defined as \cite{Komar}
\be
\textbf{J}_K=2\nabla^\nu\nabla_{[\mu}\xi_{\nu]}dx^\mu
=-\hat{\delta} d\boldsymbol{\xi}
\, , \label{KomCurr}
\ee
In accordance with the identity $\hat{\delta}^2=0$, it is easy to verify that
the Hodge coderivative of $\textbf{J}_K$ identically vanishes
($\hat{\delta}\textbf{J}_K=0$ or $\nabla_\mu J^\mu_K=0$), independently of
whether the metric tensor is on-shell or off-shell, even though $\xi^{\mu}$
is an arbitrary vector.

As what is shown in Eq. (\ref{KomCurr}), the ordinary Komar conserved current
can be completely constructed from the action of the operator
$\hat{\delta} d$ on the Killing vector $\boldsymbol{\xi}$. Apart from this
operator, one may wonder whether the other two second-order operations $\Box$ and
$d\hat{\delta}$ that also preserve the form degree could enter into the
definition of $\textbf{J}_K$. Indeed, according to the property for the Killing
vector
\be
2\Box\boldsymbol{\xi}-\chi d\hat{\delta}\boldsymbol{\xi}
-\hat{\delta}d\boldsymbol{\xi}=0
\, , \label{DivKvec}
\ee
together with the null divergence $\hat{\delta}\boldsymbol{\xi}=0$, it seems that a quite
natural way to put the usual Komar current $\textbf{J}_K$ for the Killing vector
into a novel form is to reexpress $\textbf{J}_K$ as the
linear combination of the three second-order derivative 1-forms
$\Box \boldsymbol{\xi}$, $d\hat{\delta}\boldsymbol{\xi}$
and $\hat{\delta}d\boldsymbol{\xi}$. Doing so yields the identically conserved
current
\be
\tilde{\textbf{J}}_K
=2a_1\Box \boldsymbol{\xi}-\chi d\hat{\delta}\boldsymbol{\xi}
+a_2\hat{\delta}d\boldsymbol{\xi}
\, . \label{Komcuuf2}
\ee
Here and in what follows the coefficients $\chi$, $a_1$ and $a_2$ denote
arbitrary constant parameters, but the latter two ones are constrained by
$a_1+a_2=-1$ in order to make Eq. (\ref{Komcuuf2}) coincide with the
conventional form $\textbf{J}_K=-\hat{\delta} d\boldsymbol{\xi}$
for the Killing vector. In addition to this, both the coefficients $a_1$ and $a_2$ can
be chosen for agreement with the mass of the Schwarzschild black hole when
$\tilde{\textbf{J}}_K$ is applied to define conserved charges like the Komar
integral. For another special situation where $a_1=\chi=0$, the co-closed
1-form $\tilde{\textbf{J}}_K$ is able to be adopted as the conserved current
corresponding to any vector field, such as the almost- and conformal Killing
vectors. As a result, here we arrive at the aforementioned conclusion that
each vector field can yield a conserved current. What is more, within the
framework of four-dimensional general relativity, by means of Einstein's
gravitational field equation
$R_{\mu\nu}-R g_{\mu\nu}/2=8\pi G T_{\mu\nu}$ together with
Eq. (\ref{KVRrelas}), the Komar current (\ref{Komcuuf2}) coincides with the
one presented in the work \cite{RodWai}.

The formula (\ref{Komcuuf2}) indicates that the conventional Komar conserved
current is merely built out of the second-derivative terms of one Killing
vector. However, more generally, if assumed that the conserved current
comprises at most but not just the second-order derivatives of the Killing
vector and the linearity in this vector is abandoned so that it
is permitted to appear many times in each term, we are interested in finding
the possible generalization of the Komar current (\ref{Komcuuf2}).

To find out a satisfactory prescription under the above mentioned postulate,
it is of great convenience to treat both the current and the Killing vector
field as 1-forms. As what has been demonstrated in the appendix \ref{appendA},
$\Box$, $d\hat{\delta}$ and $\hat{\delta}d$ are all the three elementary
second-order covariant derivative operators preserving the form degree
of an arbitrary differential form. Together with the help of Eq. (\ref{JV2form}),
thence one finds that the aimed conserved current $\hat{\textbf{J}}_K$ has to
take the general form
\be
\hat{\textbf{J}}_K=\tilde{\textbf{J}}_K+\bar{\textbf{J}}_K
+\lambda\boldsymbol{\xi}
\, , \label{GenKoC}
\ee
with the components of the 1-form $\bar{\textbf{J}}_K$ presented by
\be
(\bar{J}_K)_\mu=\lambda_1h_1\nabla_\mu\xi^2
+\lambda_2h_2\xi_\nu(\nabla^\rho\xi^\nu)\nabla_\rho\xi_\mu
+\lambda_3h_3\nabla^\nu\nabla_\mu\xi_\nu+\bar{U}\xi_\mu
\, . \label{JbarK}
\ee
Here the scalar $\bar{U}$ is given by
\be
\bar{U}=\lambda_4h_4 \xi^\sigma\Box\xi_\sigma
+\lambda_5h_5(\nabla^\nu\xi^2)\nabla_\nu\xi^2
+\lambda_6h_6 (\nabla_\rho\xi_\sigma)\nabla^\rho\xi^\sigma
+\lambda_7h_7R
\, . \label{Ubarxi}
\ee
In Eqs. (\ref{GenKoC}), (\ref{JbarK}) and (\ref{Ubarxi}), $\lambda=h(\xi^2)$,
$\xi^2=\xi^\sigma\xi_\sigma$, $h_i=h_i(\xi^2)$ and $\lambda_i$'s
($1\leq i\leq7$) are restricted to constant parameters in response to the
assumption that $\hat{\textbf{J}}_K$ is solely associated to the variables
$g_{\mu\nu}$ ($g^{\mu\nu}$) and $\xi^\mu$. By virtue of
the properties for Killing vectors, which are presented in the appendix
\ref{appendC}, the codifferential of the 1-form $\bar{\textbf{J}}_K$ is
read off as
\be
\hat{\delta}\bar{\textbf{J}}_K=
\lambda_1\nabla^\mu \big(h_1\nabla_\mu\xi^2\big)
+\frac{1}{2}\Big(\lambda_2h_2 +2\lambda_3\frac{dh_3}{d(\xi^2)}\Big)
R_{\rho\sigma}\xi^\rho\nabla^\sigma\xi^2
\, . \label{divJbarK}
\ee
Specially, for a certain Killing vector fulfilling $\xi^2=Const$, it is able to
ensure that $\hat{\delta}\bar{\textbf{J}}_K=0$. Nevertheless, in order to guarantee
that the divergence of the current $\hat{\textbf{J}}_K$ identically vanishes for
any Killing vector within an arbitrary spacetime manifold, in addition to that
$\hat{\delta}\tilde{\textbf{J}}_K=0$  and $\xi^\mu\nabla_\mu\lambda=0$
hold identically, $\bar{\textbf{J}}_K$ is necessarily divergenceless. This
demands that
\be
\lambda_1=0 \, , \quad
\lambda_3\frac{dh_3}{d(\xi^2)}=-\frac{1}{2}\lambda_2h_2
\, . \label{laiconst}
\ee
In Eq. (\ref{laiconst}), for example, when $\lambda_3\neq0$ and
\be
h_2=\sum_{i=0} a_i(\xi^2)^i
\, , \label{h2exprs}
\ee
the solutions for the second equation are
\be
h_3=-\frac{\lambda_2}{2\lambda_3}\sum_{i=0}\frac{a_i}{i+1}(\xi^2)^{i+1}+C
\, . \label{h6expr}
\ee
Here and henceforth, the quantity $C$, together with all the coefficients
$a_i$'s, stands for arbitrary constant parameters. More generally, in the
absence of the requirement that $\lambda$ is dependent of at most
second-order derivatives, one interesting solution
for the constraint equation $\xi^\mu\nabla_\mu\lambda=0$ is the one
\be
\lambda=\sum_i \big(a_i F_i+b_i W_i\big)
+ \sum_{i,j} k_{ij} \tilde{F}_i\tilde{W}_j
\, . \label{Sollam}
\ee
Here and in what follows the coefficients $b_i$'s and $k_{ij}$'s denote
constant parameters. The scalars $\{F_i,i=0,1,\cdot\cdot\cdot\}$ and
$\{W_i,i=0,1,\cdot\cdot\cdot\}$ in Eq. (\ref{Sollam}) are defined through
\bea
F_i&=&f_i^0(g)f_i^1(\mathcal{R})f_i^2(\nabla \mathcal{R})
f_i^3(\nabla\nabla \mathcal{R})
f_i^4(\nabla\nabla\nabla \mathcal{R})\cdot\cdot\cdot
 \, , \nn \\
W_i&=&w_i^0(g)w_i^1(\xi)w_i^2(\nabla\xi)w_i^3(\nabla\nabla \xi)
w_i^4(\nabla\nabla\nabla \xi)\cdot\cdot\cdot
\, ,  \label{defW}
\eea
while the scalars $\tilde{F}_i$'s and $\tilde{W}_i$'s take the forms similar to
$F_i$'s and $W_i$'s respectively. Notice that $\mathcal{R}$ in Eq. (\ref{defW})
denotes the standard Riemann tensor $R_{\rho\sigma\mu\nu}$ and $f_i^k(X)=1$
($w_i^k(X)=1$) when such a term is absent, so it could be set $F_0=W_0=1$.
For instance, assumed that $i=1$, the Ricci scalar $R$ could be written as
$R=f_1^0f_1^1$ with $f_1^0(g)=g^{\rho\mu}g^{\sigma\nu}$ and
$f_1^1(\mathcal{R})=R_{\rho\sigma\mu\nu}$. What is
more, if $i=3$, the $\xi^\sigma\Box\xi_\sigma$ part in the $\lambda_4$ term
of Eq. (\ref{JbarK}) could be expressed as $w_3^0w_3^1w_3^3$, where
$w_3^0=g_{\rho\sigma}g^{\mu\nu}$, $w_3^1=\xi^\rho$ and
$w_3^3=\nabla_\mu\nabla_\nu \xi^\sigma$. According to Eq. (\ref{CdW}), it
is shown that the Lie derivatives of $F_i$'s and $W_i$'s along the Killing
vector $\xi^\mu$ disappear, giving rise to that $\xi^\mu\nabla_\mu\lambda=0$
holds identically.

Therefore, we are able to put forward a generalized conserved current
$\check{\textbf{J}}_K$ for any Killing vector, irrespective of whether
the metric is on-shell or off-shell.
$\check{\textbf{J}}_K$ depends at most upon the second-order derivatives
of the Killing vector $\xi^\mu$ and is expressed as
\be
\check{\textbf{J}}_K=\tilde{\textbf{J}}_K
+h(\xi^2)\boldsymbol{\xi}
+\bar{\textbf{J}}_K\big|_{\lambda_1=0}
\, . \label{ckJK}
\ee
Here it is worth noting that the 1-form $\bar{\textbf{J}}_K$ is constrained
by Eq. (\ref{laiconst}). The current $\check{\textbf{J}}_K$ mainly differs from the
usual Komar current $\tilde{\textbf{J}}_K$ by non-linear terms of the Killing
vector. As a direct result of $\check{\textbf{J}}_K$, it
is allowed to single out the conserved current
$\tilde{\textbf{J}}_K+R\boldsymbol{\xi} =2G_{\mu\nu}\xi^\nu dx^\mu$
with $a_1+a_2=-1$. In fact, the quantity $\bar{U}$ can be covered by the
scalars $F_i$'s and $W_i$'s given by Eq. (\ref{defW}). Consequently, a
straightforward generalization of Eq. (\ref{ckJK}) leads to the more
general conserved current
\be
\textbf{J}_{GK}=\tilde{\textbf{J}}_K
+(\lambda+C)\boldsymbol{\xi}
\, . \label{GenJK}
\ee
Notice that $\lambda$ is presented by Eq. (\ref{Sollam}) in the above
equation and $h(\xi^2)$ in Eq. (\ref{ckJK}) has been incorporated
into $W_i$. What is more, if the spacetime is Ricci-flat,
$\hat{\textbf{J}}_K$ acquires the following form
\be
\hat{\textbf{J}}_K\big|_{R_{\mu\nu}=0}=
\hat{\textbf{J}}_K\big|_{\lambda_1=0}
\, . \label{TidJKRf}
\ee
In accordance with Eq. (\ref{ckJK}), if the number of the Killing vector in
any term of the conserved currents is demanded to be one, one can find that
$\check{\textbf{J}}_K$ differs from $\tilde{\textbf{J}}_K$ by the terms
proportional to the Killing vector, that is,
\be
\check{\textbf{J}}_K\rightarrow (a_1+a_2)\hat{\delta}d\boldsymbol{\xi}
+(b_1+b_2 R)\boldsymbol{\xi}
\, . \label{chJsp2}
\ee
As a result, regardless of the terms $b_1\boldsymbol{\xi}$
and $b_2 R\boldsymbol{\xi}$, under the condition that the conserved
current is permitted to rely at most upon second-order derivatives of a
Killing vector, such a current has to take the form proportional to
the co-closed 1-form $\hat{\delta}d\boldsymbol{\xi}$. To this point, it
could be concluded that the ordinary Komar conserved current is ``unique".

For instance, in the framework of the Einstein gravity theory described by
the Einstein-Hilbert Lagrangian
\be
 L_{gr}=\sqrt{-g}(R-2\Lambda)
 \, , \label{EHLag}
\ee
where $\Lambda$ is the cosmological constant, according to Eq. (\ref{chJsp2}),
a conserved current can be proposed as
\be
\textbf{J}_{EH}=-\hat{\delta}d\boldsymbol{\xi}
+\frac{2nC\Lambda}{n-2}\boldsymbol{\xi}
\, . \label{CurrEH}
\ee

\subsection{The extended Komar currents associated to generic
vector fields}\label{ss22}

We move on to consider the generalization for the Komar current (\ref{Komcuuf2}) to the one
with respect to an arbitrary vector $V^\mu$ in $n$-dimensional spacetime.
For the sake of convenience, the symmetrization and anti-symmetrization of the covariant
derivative of this vector are supposed to take the respective forms
\be
2\nabla_{(\mu} V_{\nu)}=\Phi_{\mu\nu} \, , \quad
2\nabla_{[\mu} V_{\nu]}=\Psi_{\mu\nu}
\, . \label{GenVecEq}
\ee
Obviously, the tensor $\Phi_{\mu\nu}$ ($\Psi_{\mu\nu}$) is (anti-)symmetric
under the interchange of the two spacetime indices. The contraction $\Phi$
between $\Phi_{\mu\nu}$ and the
metric tensor is denoted by $\Phi=g^{\mu\nu}\Phi_{\mu\nu}=2\nabla^\mu V_\mu$.
In accordance with the notation of differential forms, the divergence of $\Phi_{\mu\nu}$
is given by
\be
\nabla^\mu \Phi_{\mu\nu}dx^\nu=2\Box\textbf{V}-\hat{\delta}d \textbf{V}
\, , \label{DivPh}
\ee
with the help of Eq. (\ref{ComBhdd}), while the divergence of
$\Psi_{\mu\nu}$ is
$\nabla^\mu \Psi_{\mu\nu}dx^\nu=\hat{\delta}d \textbf{V}$.
One can check that the
first-order covariant derivatives of $V^\mu$ can be completely expressed
through the three quantities $\Phi$, $\Phi_{\mu\nu}$ and $\Psi_{\mu\nu}$
attributed to the decomposition relations
$2\nabla_\mu V_\nu=\Phi_{\mu\nu}+\Psi_{\mu\nu}$ and
$2\nabla_\nu V_\mu=\Phi_{\mu\nu}-\Psi_{\mu\nu}$.
According to this, it will be seen below that they are potential
candidates to play the roles of the basic ingredients in the construction
for the conserved current with respect to the vector $V^\mu$.

As the matter of fact, inspired by the form (\ref{Komcuuf2}) for the usual
Komar current, we assume that
a conserved current $\textbf{J}_{V}$ ($\hat{\delta}\textbf{J}_{V}=0$) associated
with the vector $V^\mu$ is a linear combination of the vectors $X_{(i)}^\mu$'s
with weights $k_i$'s, namely,
\be
J_{V}^\mu=\sum_{i=1} k_i X_{(i)}^\mu
\, . \label{JVgenform}
\ee
Throughout the present work, the scalars $k_i$'s stand for arbitrary constant
parameters. As a straightforward extension of the conventional Komar current,
it is natural to impose the requirements that the number of $V^\mu$ in each
$X_{(i)}^\mu$ is restricted to one and $X_{(i)}^\mu$'s are all the elements
that can generate an arbitrary second-order derivative vector field of
$V^\mu$. Hence the former ensures that the conserved current is linear in the
vector field. As a consequence of Eq. (\ref{Th2npF}), $X_{(i)}^\mu$'s have
to be the quantities $\nabla^\mu\Phi_{\mu\nu}$, $\nabla_\nu\Phi$ and
$\nabla^\mu\Psi_{\mu\nu}$. This implies that the conserved current
$\textbf{J}_{V}$ can be further expressed as the following general form:
\bea
\textbf{J}_{V}&=&k_1\big(\nabla^\mu \Phi_{\mu\nu}\big) dx^\nu+k_2 d\Phi
+(k_1+k_3) \big(\nabla^\mu \Psi_{\mu\nu}\big) dx^\nu    \nn \\
&=& 2k_1\Box \textbf{V}+2k_2 d\hat{\delta} \textbf{V}
+k_3 \hat{\delta}d \textbf{V}
\, , \label{CurrofGV}
\eea
as long as $\textbf{V}$ obeys the constraint
\be
k_1\nabla^\mu\nabla^\nu \Phi_{\mu\nu}+k_2\Box\Phi =0
\, . \label{divofCofGV}
\ee
Apparently, two simple cases of this constraint are the ones in which $k_1=0=k_2$ for
arbitrary vectors and $\Phi_{\mu\nu}=0$ for the Killing vectors.
Noteworthily, apart from the form (\ref{divofCofGV}), by means of Eq. (\ref{CoBoCRV}),
we are able to reformulate the constraint condition as the following form
\be
2(k_1+k_2)\Box\Phi
+k_1(V^\mu\nabla_\mu R
+R_{\mu\nu}\Phi^{\mu\nu})=0
\, . \label{divofCofGV2}
\ee
Eq. (\ref{divofCofGV2}) directly leads to the conclusion that the 1-form
$\textbf{J}_{V}$ is co-closed for any vector $V^\mu$ fulfilling the
condition $\Box\nabla_\mu V^\mu=0$ when the spacetime is Ricci-flat. Enlightened by
the structure of $\textbf{J}_{V}$, it is reasonable for us to introduce the most
general quadratic operation $\mathrm{P}(k_1,k_2,k_3)$ through the linear
combination of the operators $\Box$, $d\hat{\delta}$ and $\hat{\delta}d$,
which automatically leaves the form degree of any differential form unchanged
and is expressed as
\be
\mathrm{P}(k_1,k_2,k_3)=k_1\Box +k_2 d\hat{\delta}+k_3 \hat{\delta}d
\, . \label{OpefBhdd}
\ee
By making use of such an operation, we are able to rewrite the current
$\textbf{J}_{V}$ as $\textbf{J}_{V}=\mathrm{P}(2k_1,2k_2,k_3)\textbf{V}$.
What is more, if both $k_1$ and $k_2$ are permitted to be functions of
coordinates, it is worth noting that the condition (\ref{divofCofGV})
should be modified as
\be
(\nabla^\mu k_1)(\nabla^\nu \Phi_{\mu\nu}-\nabla^\nu \Psi_{\mu\nu})
+k_1\nabla^\mu\nabla^\nu \Phi_{\mu\nu}
+k_2\Box\Phi+(\nabla^\mu k_2)\nabla_\mu\Phi=0
\, . \label{divofCk1k2}
\ee

According to the second equality in Eq. (\ref{CurrofGV}), it is observed
that $\textbf{J}_{V}$ covers the usual Komar current $\textbf{J}_{K}$
as the simplest case since $\Phi_{\mu\nu}=0$ for the Killing vector
$\textbf{V}=\boldsymbol{\xi}$. More generally, when the vector $V^\mu$ is
the so-called homothetic vector field (or homothety), namely,
$\Phi_{\mu\nu}=C g_{\mu\nu}$ with the arbitrary constant factor $C$, the current
$\textbf{J}_{V}$ turns to $\textbf{J}_{V}=(k_1+k_3)\hat{\delta}d\textbf{V}$.
For another situation
where the spacetime is Ricci-flat, namely, $R_{\mu\nu}=0$, Eq. (\ref{PoppForm})
enables us to obtain the current $\textbf{J}_{V}=(2k_1+k_3)\hat{\delta}d\textbf{V}$
for arbitrary $V^\mu$ when $k_2=-k_1$. In fact, such a case can be covered by
the one with $k_1=k_2=0$.

Lastly, motivated by Eq. (\ref{DivKvec}), we take into consideration of the
current corresponding to the vector obeying the equation
\be
\mathrm{P}(2,\chi,\lambda) \textbf{V}=\textbf{Y}(\textbf{V})
\, , \label{SpecV}
\ee
where $\textbf{Y}(\textbf{V})$ is an arbitrary 1-form with respect to the
vector $V^\mu$ and the metric. In some sense, Eq. (\ref{SpecV}) can be
regarded as the extension of Eq. (\ref{DivKvec}) for the Killing vector
field or the more general semi-Killing vector field \cite{sKVK} defined
through the two constraints $2\Box\textbf{V}=\hat{\delta}d\textbf{V}$ and
$\hat{\delta}\textbf{V}=0$, and it will be shown below that this equation
also covers the ones associated to the almost- and conformal Killing
vectors. As a specific example for Eq. (\ref{SpecV}), here we present the
equation $\mathrm{P}(2,0,-1) \textbf{V}=\kappa \textbf{V}$ with a scalar
$\kappa$. It has been demonstrated that an approximate Killing vector
field might be defined via solving an eigenvalue problem associated
to this equation \cite{MaASSG,AppKB}. In a rather simple situation where
$\textbf{Y}=0$ and $\chi=\lambda=-2$, Eq. (\ref{PoppForm}) enables us to
simplify Eq. (\ref{SpecV}) as $R_{\mu\nu} V^\nu=0$. What is more,
according to  Eq. (\ref{SpecV}), one may propose another generalization
to the Killing vector field $\boldsymbol{\zeta}$ in the absence of the
requirement to preserve the linearity in the vector field, defined through
\be
\mathrm{P}(2,\chi,-1) \boldsymbol{\zeta}=
\lambda_1\mathcal{L}_\zeta \boldsymbol{\zeta}
+\lambda_2\boldsymbol{\zeta}\hat{\delta} \boldsymbol{\zeta}
\, . \label{GeKVzet}
\ee
Here $\mathcal{L}_\zeta \boldsymbol{\zeta}
=2\zeta^\nu\nabla_{(\mu}\zeta_{\nu)} dx^\mu$
and $\mathcal{L}_\zeta$ denotes the Lie derivative along the vector $\zeta$.

When $\textbf{Y}=0$ in Eq. (\ref{SpecV}), the conserved current
(\ref{CurrofGV}) for the vector determined by this equation transforms into
$\textbf{J}_{V}=(2k_2-\chi k_1) d\hat{\delta} \textbf{V}+
(k_3-\lambda k_1)\hat{\delta}d\textbf{V}$, while the constraint
(\ref{divofCofGV}) becomes $(2k_2-\chi k_1)\Box\Phi=0$. Specially,
when $\chi=2k_2/k_1$, the resulting conserved current further becomes
$\textbf{J}_{V}=(k_3-\lambda k_1)\hat{\delta}d\textbf{V}$. Apart from
$\textbf{J}_{V}$, the current associated with such a vector field
could be defined as
\bea
\textbf{J}^Y_{V}&=& a_1(2\Box \textbf{V}+\chi d\hat{\delta} \textbf{V}
-\textbf{Y})+a_2 \hat{\delta}d \textbf{V} \nn \\
&=&(a_2-\lambda a_1) \hat{\delta}d \textbf{V}
\, . \label{CCuofYV}
\eea
It is apparently divergence-free and recovered by $\textbf{J}_{V}$. In parallel,
considering the vector field satisfying the well-known Proca equation
$\mathrm{P}(0,0,1)\textbf{V}=\lambda \textbf{V}$
(the constant parameter $\lambda\neq0$) or the more general one
$\mathrm{P}(0,\chi,\lambda) \textbf{V}=\textbf{Y}(\textbf{V})$, we obtain
the related identically conserved current
$\textbf{J}^P_{V}=a_1\textbf{V}+a_2\hat{\delta}d \textbf{V}$
or $\tilde{\textbf{J}}^Y_{V}= a_1(\chi d\hat{\delta} \textbf{V}
-\textbf{Y})+a_2 \hat{\delta}d \textbf{V}$.

\subsection{Constructing conserved currents out of $\textbf{J}_{V}$}
\label{ss23}

In order to illustrate the generalized Komar current $\textbf{J}_{V}$, within
this subsection, we shall take into consideration of a couple of examples
associated to several interesting specific vector fields. Particularly, we
are going
to concentrate on the vectors describing approximate symmetries, such as the
almost- and conformal Killing vectors, together with the affine collineation
vector, attributed to the fact that Killing vector fields corresponding to
exact symmetries can not be generally found in many physically important
spacetimes.

First, when $\Phi_{\mu\nu}$ in Eq. (\ref{GenVecEq}) is restricted to
\be
\nabla^\mu\Phi_{\mu\nu}=\frac{1}{2}\chi \nabla_\nu\Phi
\, , \label{almKconPh}
\ee
or equivalently $2\nabla^\mu\nabla_{(\mu} V_{\nu)}=\chi\nabla_\nu\nabla_{\mu} V^{\mu}$,
the vector $V^\mu$ coincides with the almost-Killing vector $\kappa^\mu$,
defined through \cite{CHTau,BoCaPaL,RuPaBo}
\be
\Box\kappa_\mu+R_\mu^\sigma \kappa_\sigma
=(\chi-1)\nabla_\mu(\nabla\cdot\kappa)
\, . \label{DiveraKV}
\ee
Here $\nabla\cdot\kappa=\nabla_\mu\kappa^\mu=\hat{\delta}\boldsymbol{\kappa}$.
The above equation can be regarded as the generalization of the divergence
for the ordinary Killing equation (the special case where $\chi=1$) and
it is reexpressed as
\be
\mathrm{P}(2,-\chi,-1) \boldsymbol{\kappa}=0
\,  \label{AKEinDform}
\ee
by virtue of Eq. (\ref{PoppForm}). It corresponds to the $\lambda=-1$ case
of Eq. (\ref{SpecV}) and completely coincides with Eq. (\ref{DivKvec}) for
the Killing vector field under the condition that
$\hat{\delta}\boldsymbol{\kappa}=0$. Provided that the almost-Killing vector
fulfills $\hat{\delta}d\hat{\delta}\boldsymbol{\kappa}=0$, we have
$\hat{\delta}\Box\boldsymbol{\kappa}=0=\Box\hat{\delta}\boldsymbol{\kappa}$
and $\hat{\delta}\mathbf{\Omega}(\boldsymbol{\kappa})=0$, where $\mathbf{\Omega}$
is given by Eq. (\ref{DefOmega}). What is more, when $V^\mu=\kappa^\mu$, the
current $\textbf{J}_{V}$ goes to
\be
\textbf{J}_{AKV}=(\chi k_1+2k_2)d\hat{\delta}\boldsymbol{\kappa}
+(k_1+k_3)\hat{\delta}d\boldsymbol{\kappa}
\, . \label{CurraKV}
\ee
Here we have to impose the constraint $k_1\chi+2k_2=0$ or
$\hat{\delta}d\hat{\delta}\boldsymbol{\kappa}=0$ to guarantee that
$\hat{\delta}\textbf{J}_{AKV}=0$. Particularly, when $k_1=1$,
$k_2=-\chi/2$ and $k_3=-2$, it covers the current $J^\nu_{AK}$ in
\cite{JuFeng,RuPaBo}. To see this clearly, note that Eq. (\ref{CoBoCRV})
is of great use.

Second, if $V^\mu$ is a conformal Killing vector, which maintains the metric
up to an arbitrary multiplicative factor $\phi$ (the so-called conformal
factor) and is defined by
\be
\Phi_{\mu\nu}=\phi g_{\mu\nu}
\, , \label{ConKVdef}
\ee
by analogy with the situation for the almost-Killing vector,
Eq. (\ref{SpecV}) becomes
\be
\mathrm{P}(2,-2/n,-1)\textbf{V}=0
\, \label{SpecVCKV}
\ee
and the current $\textbf{J}_{V}$ transforms into
\be
\textbf{J}_{CKV}=(k_1+nk_2)d\phi+(k_1+k_3) \hat{\delta}d \textbf{V}
\, , \label{CCforConV}
\ee
provided that $k_1+nk_2=0$ or $\Box\phi =0$. Obviously, Eq. (\ref{SpecVCKV})
for the conformal Killing vector is just a special case of Eq. (\ref{SpecV})
for the almost-Killing vector. Letting $k_1=1$, $k_2=-1/4$ and
$k_3=-2$ in the above equation, one finds that $J^\mu_{CKV}$ is equivalent
to the four-dimensional current $J^\mu_{C}$ associated to a conformal vector
field in \cite{JuFeng}.

Third, for the affine collineation vectors satisfying
$\nabla_\rho \Phi_{\mu\nu}=0$, we gain
$2\Box\textbf{V}=\hat{\delta}d \textbf{V}$ and $d\Phi=0$. Thus the corresponding
current $\textbf{J}_{ACV}=(k_1+k_3) \hat{\delta}d \textbf{V}$, taking the
same form as the one with respect to the Killing vector.

Fourth, we take into account the conserved current for the vector
being of the form
\be
\Phi_{\mu\nu}=2B_{\mu\nu}=(2\nabla^\rho\nabla^\sigma
+R^{\rho\sigma})C_{\mu\rho\nu\sigma}
\, , \label{VofBC}
\ee
where $C_{\mu\rho\nu\sigma}$ is the standard Weyl tensor, while $B_{\mu\nu}$
denotes the traceless, symmetric and conserved Bach tensor, corresponding to
$g^{\mu\nu}B_{\mu\nu}=0$, $B_{\mu\nu}=B_{\nu\mu}$ and
$\nabla^\nu B_{\mu\nu}=0$ respectively. For such a vector, the conserved current
$\textbf{J}_{BV}=(k_1+k_3) \hat{\delta}d \textbf{V}$. Besides, for the Killing
vector $\xi^\mu$, another conserved current associated to the Bach tensor is
$J_B^\mu=k_1B^{\mu\nu}\xi_\nu$.

Fifth, supposed that the vector $V^\mu$ can be expressed as the Hodge coderivative
of some 2-form $\boldsymbol{\omega}$ in analogy with the works \cite{KasKinHDG,BaZygrg},
that is, $\textbf{V}=\hat{\delta}\boldsymbol{\omega}$ (here the divergenceless vector
field $V^\mu$ is unnecessarily restricted to the Killing vector), the conserved current
$\textbf{J}_{V}$ then turns into
\be
\textbf{J}_{\boldsymbol{\omega}}=2k_1
\big(R^\nu_\mu\nabla^\rho\omega_{\rho\nu}\big) dx^\mu
+(2k_1+k_3) \hat{\delta}d \hat{\delta}\boldsymbol{\omega}
\, , \label{JVome}
\ee
with the help of Eq. (\ref{PoppForm}). Here
$\hat{\delta}\textbf{J}_{\boldsymbol{\omega}}=0$ holds identically only if
$\nabla^\mu\big(R_{\mu\nu}\nabla_\rho\omega^{\rho\nu}\big)=0$. Particularly,
in the framework of the $n$-dimensional Einstein gravity theory described
by the conventional Einstein-Hilbert Lagrangian (\ref{EHLag}), by means of the
equation of motion $R_{\mu\nu}=2\Lambda g_{\mu\nu}/(n-2)$, we obtain the
conserved current
\be
\textbf{J}^{gr}_{\boldsymbol{\omega}}=\frac{4k_1}{n-2}
\Lambda\hat{\delta}\boldsymbol{\omega}
+(2k_1+k_3) \hat{\delta}d\hat{\delta}\boldsymbol{\omega}
\, . \label{JVomexi}
\ee
When $V^\mu$ is just the Killing vector $\xi^\mu$, namely,
$\boldsymbol{\xi}=\hat{\delta}\boldsymbol{\omega}$,
the co-closed 1-form
$\textbf{J}^{gr}_{\boldsymbol{\omega}}\mid_{2k_1=-1,k_3=0}$
is consistent with the conserved current (\ref{CurrEH}), as well as the one
given in \cite{KasKinHDG,BaZygrg}. It has been demonstrated in \cite{KaRTinJ}
that $\textbf{J}^{gr}_{\boldsymbol{\omega}}\mid_{2k_1=-1,k_3=0}$ might provide a
novel understanding on the thermodynamics of asymptotically anti-de Sitter
(AdS) black holes. Besides, if $V^\mu$ is the solution of the Proca equation
$\hat{\delta}d\textbf{V}=\lambda \textbf{V}$, yielding
$\hat{\delta}\textbf{V}=0$, the 1-form field
$\textbf{J}^{gr}_{\boldsymbol{\omega}}$ can be proposed as its corresponding
conserved current.

\section{Various generalizations of the current $\textbf{J}_{V}$}\label{three}

In the present section, we are going to take into consideration of the generalizations of
the conserved current $\textbf{J}_{V}$. We shall mainly concentrate on the extended current
that is dependent of at most second-order derivatives of the vector field, as well as the one
encompassing higher-order derivative terms of the vector.

First, if it is allowed that the current is able to admit the
first-order covariant derivatives of $V^\mu$, the current $\textbf{J}_{V}$
may be generalized to the one
\be
\tilde{\textbf{J}}_{V}=\textbf{J}_{V}+\textbf{J}_{V1}
\, , \label{TildJV}
\ee
in which the 1-form $\textbf{J}_{V1}$ is presented by
\bea
(\textbf{J}_{V1})_\mu&=&\big(b_1\Phi g_{\mu\nu}
+(b_2+b_3)\Phi_{\mu\nu}+b_3 \Psi_{\mu\nu}\big)V^\nu \nn \\
&=&b_1\Phi V_\mu
+b_2\Phi_{\mu\nu}V^\nu+b_3 \nabla_\mu(V\cdot V)
\, . \label{DefofJV1}
\eea
Here $b_i$ ($i=1,2,3$) are constant parameters or functions of
the scalar $V\cdot V$. To make the coderivative of $\tilde{\textbf{J}}_{V}$
vanish, the vector $V^\mu$ has to obey that
$\hat{\delta}\textbf{J}_{V1}=-\hat{\delta}\textbf{J}_{V}$.
Besides, if the number of the vector $V^\mu$ in each term is not restricted
to one, from a mathematical point of view, the most general 1-form that
contains the terms of no higher than second order covariant derivative of
the vector $V^\mu$ is
\be
\check{\textbf{J}}_{V}=\textbf{J}_{V}+\textbf{J}_{V1}
+\textbf{J}_{V2}+\lambda\textbf{V}
\, , \label{JV12Coder}
\ee
where $\textbf{J}_{V2}$ is presented by Eq. (\ref{JV2form}) in the appendix
\ref{appendB}. $\check{\textbf{J}}_{V}$ is the conserved current associated with the
vector satisfying the equation $\hat{\delta}\check{\textbf{J}}_{V}=0$.

In some sense, the generalized current $\check{\textbf{J}}_{V}$ given by
Eq. (\ref{JV12Coder}) can be applied to understand the Noether current
corresponding to the diffeomorphism symmetry of a spacetime manifold, generated
by an arbitrary vector field $V^\mu$, in the context of the Einstein
gravity equipped with the Lagrangian (\ref{EHLag}). According to the standard
Noether method, the conserved current associated with the diffeomorphism
symmetry is given by $\textbf{J}_{NC}=-\hat{\delta}d\textbf{V}$
\cite{IyerWald,PengN}. On the other hand, as what has been mentioned above,
supposed that the current depends solely
on the terms proportional to at most the second-order derivatives of a vector, the
satisfactory conserved current for $V^\mu$ has to be $\check{\textbf{J}}_{V}$. However, since
$\check{\textbf{J}}_{V}$ is necessarily divergence-free for arbitrary vectors, it is demanded
that the 1-forms $\textbf{J}_{V1}$ and $\textbf{J}_{V2}$, as well as the parameter
$\lambda$ together with the ones $k_1$, $k_2$ in $\textbf{J}_{V}$, disappear.
This indicates that the current $\check{\textbf{J}}_{V}$ further becomes
$\check{\textbf{J}}_{V}=k_3\hat{\delta}d\textbf{V}$. In other words, the co-closed
1-form $k_3\hat{\delta}d\textbf{V}$, which maintains the form degree of the
arbitrary 1-form $\textbf{V}$ and has the lowest differential order, is the
only covariant expression of the conserved current that is identically conserved
and linear in any vector field. To this point, one is able
to observe that the Noether current $\textbf{J}_{NC}$ has to contain the expression
proportional to $\hat{\delta}d\textbf{V}$. The above discussions maybe give an
explanation why Komar could succeed in finding out the simple expression
$2\nabla_\nu\nabla^{[\mu}\xi^{\nu]}$ (here $\xi^\mu$ is an arbitrary vector
field according to the notation in \cite{Komar}) for the conserved current
in \cite{Komar} without the assistance of the standard Nother approach.

Next, if the conserved current, which is required to be linear in the vector
$V^\mu$, is permitted to consist of all the terms proportional to $(2i)$-th-order
(the integer $i$ is restricted to $1\leq i\leq N$ for some given positive
integer $N$) derivatives of this vector, the current $\textbf{J}_{V}$ can be
straightforwardly generalized to the higher-derivative one
$\mathbf{\mathcal{J}}_{HV}^{(2N)}$, taking the following general form
\bea
\mathbf{\mathcal{J}}_{HV}^{(2N)}&=&\sum_{i=1}^N
\bar{\mathbf{\mathcal{J}}}_{HV}^{(2i)} \, , \nn \\
\bar{\mathbf{\mathcal{J}}}_{HV}^{(2i)}&=& \sum_{l}\alpha_{il}
\mathcal{O}_{il}\big(\Box,d,\hat{\delta}\big)\textbf{V}
\, . \label{CalJ2Ndef}
\eea
In Eq. (\ref{CalJ2Ndef}), $\alpha_{il}$'s are constant
parameters, and $l$ denotes an arbitrary non-zero permutation of the
total $(2i-K)$ operators $\Box$, $d$ and $\hat{\delta}$, where the integer
$K$ ($0\leq K\leq i$) represents the number of $\Box$ in each
operator $\mathcal{O}_{il}$.
The number of $\hat{\delta}$ or $d$ is $(i-K)$ because the coderivative
$\hat{\delta}$ has to be paired with the exterior derivative $d$. Besides,
the combinations $d^k$ and $\hat{\delta}^k$ ($k\geq 2$) are not allowed to
exist due to the fact that $\hat{\delta}^2=d^2=0$. Generally, the 1-forms
$\mathcal{O}_{il}\textbf{V}$ have the forms
$\Box^i\textbf{V}$, $\big(d\hat{\delta}\big)^i\textbf{V}$ and
$\big(\hat{\delta}d\big)^i\textbf{V}$, together with the following structures
mixing the $(\Box,\hat{\delta},d)$ operators
\bea
\Box\cdot\cdot\cdot \hat{\delta}\cdot\cdot\cdot d \cdot\cdot\cdot
\textbf{V} \, , \nn \\
d\cdot\cdot\cdot\Box \cdot\cdot\cdot\hat{\delta}\cdot\cdot\cdot
\textbf{V} \, , \nn \\
\hat{\delta}\cdot\cdot\cdot\Box \cdot\cdot\cdot d \cdot\cdot\cdot
\textbf{V}
\, . \label{Ojtype}
\eea
For instance, when $i=1,2$, the 1-forms
$\bar{\mathbf{\mathcal{J}}}_{HV}^{(2)}$ and
$\bar{\mathbf{\mathcal{J}}}_{HV}^{(4)}$,
consisting of second-order and fourth-order derivatives of the vector
$\textbf{V}$ respectively, can be expressed as
\bea
\bar{\mathbf{\mathcal{J}}}_{HV}^{(2)}
&=&\alpha_{11}\Box\textbf{V}
+\alpha_{12}d\hat{\delta}\textbf{V}
+\alpha_{13}\hat{\delta}d\textbf{V} \, , \nn \\
\bar{\mathbf{\mathcal{J}}}_{HV}^{(4)}&=&
\alpha_{21}\Box^2\textbf{V}
+\alpha_{22}\big(d\hat{\delta}\big)^2\textbf{V}
+\alpha_{23}\big(\hat{\delta}d\big)^2\textbf{V}\nn \\
&&+\alpha_{24}\Box\hat{\delta}d\textbf{V}
+\alpha_{25}\Box d\hat{\delta}\textbf{V}
+\alpha_{26}d\Box \hat{\delta}\textbf{V}  \nn \\
&&+\alpha_{27}d\hat{\delta}\Box \textbf{V}
+\alpha_{28}\hat{\delta}d\Box \textbf{V}
+\alpha_{29}\hat{\delta}\Box d\textbf{V}
\, . \label{CalbarJ4}
\eea
In the above equation, $\bar{\mathbf{\mathcal{J}}}_{HV}^{(2)}$ is completely
equivalent to $\textbf{J}_{V}$ given by Eq. (\ref{CurrofGV}). Its combination
with $\bar{\mathbf{\mathcal{J}}}_{HV}^{(4)}$ makes us write down the
expression
\be
\mathbf{\mathcal{J}}_{HV}^{(4)}=
\bar{\mathbf{\mathcal{J}}}_{HV}^{(2)}
+\bar{\mathbf{\mathcal{J}}}_{HV}^{(4)}
\, . \label{CalJ4}
\ee
For another example on the conserved currents consisting of the sixth-order
derivatives of an arbitrary Killing vector, see the work \cite{PenZL}.

As a special case of $\mathbf{\mathcal{J}}_{HV}^{(2N)}$, we consider the following
current generated by the $\big(\Box,d\hat{\delta},\hat{\delta}d\big)$ operators,
together with their various combinations
\be
\textbf{J}_{HV}^{(2N)}=\sum_{j} k_j
f_j\big(\Box,d\hat{\delta},\hat{\delta}d\big)\textbf{V}
\, . \label{HCforV0}
\ee
Concretely, $\textbf{J}_{HV}^{(2N)}$ can be written as
\be
\textbf{J}_{HV}^{(2N)}=\sum_{i=1}^{N}\Big(\bar{\textbf{J}}_{HV}^{(2i)}
+\tilde{\textbf{J}}_{HV}^{(2i)}\Big)
\, . \label{HCforV}
\ee
Here the 1-forms $\{\bar{\textbf{J}}_{HV}^{(2i)}, i=1,\cdot\cdot\cdot,N\}$,
which have to satisfy the constraint
$\hat{\delta}\sum_{i=1}^{N}\bar{\textbf{J}}_{HV}^{(2i)}=0$ in order to
guarantee the current
$\textbf{J}_{HV}^{(2N)}$ to be co-closed, is defined through
\be
\bar{\textbf{J}}_{HV}^{(2i)}
=\sum_{i_2,i_4,\cdot\cdot\cdot}\sum_{j_0,j_1,\cdot\cdot\cdot}
s_{i_2i_4\cdot\cdot\cdot}^{j_0j_1\cdot\cdot\cdot}
\mathrm{P}_1^{j_0}
\bar{\textbf{V}}_{i_2i_4\cdot\cdot\cdot}^{j_1j_2\cdot\cdot\cdot}+
\sum_{i_2,i_4,\cdot\cdot\cdot}\sum_{j_0,j_1,\cdot\cdot\cdot}
t_{i_2i_4\cdot\cdot\cdot}^{j_0j_1\cdot\cdot\cdot}
\mathrm{P}_2^{j_0}
\bar{\textbf{V}}_{i_2i_4\cdot\cdot\cdot}^{j_1j_2\cdot\cdot\cdot}
\,  \label{bJVdef}
\ee
with the 1-forms $\bar{\textbf{V}}_{i_2i_4\cdot\cdot\cdot}^{j_1j_2\cdot\cdot\cdot}$
defined by
\be
\bar{\textbf{V}}_{i_2i_4\cdot\cdot\cdot}^{j_1j_2\cdot\cdot\cdot}
=\big(\mathrm{P}_1^{j_1}\mathrm{P}_{i_2}^{j_2}
\mathrm{P}_1^{j_3}\mathrm{P}_{i_4}^{j_4}
\mathrm{P}_1^{j_5}\mathrm{P}_{i_6}^{j_6}\cdot\cdot\cdot\big)
\textbf{V}
\, , \label{baVdef}
\ee
while the co-closed 1-forms
$\{\tilde{\textbf{J}}_{HV}^{(2i)}, i=1,\cdot\cdot\cdot,N\}$ can be expressed
as a similar form like the usual Komar current $\textbf{J}_{K}$,
that is,
\be
\tilde{\textbf{J}}_{HV}^{(2i)}=-\hat{\delta}d\tilde{\textbf{V}}^{(i)}
\, , \label{tiJVdef}
\ee
by virtue of the 1-form $\tilde{\textbf{V}}^{(i)}$, given by
\be
\tilde{\textbf{V}}^{(i)}=
\sum_{i_2,i_4,\cdot\cdot\cdot}\sum_{j_0,j_1,\cdot\cdot\cdot}
u_{i_2i_4\cdot\cdot\cdot}^{j_0j_1\cdot\cdot\cdot}
\mathrm{P}_3^{j_0-1}
\bar{\textbf{V}}_{i_2i_4\cdot\cdot\cdot}^{j_1j_2\cdot\cdot\cdot}
\, . \label{tilVdef}
\ee
In Eqs. (\ref{bJVdef}) and (\ref{tilVdef}),
$s_{i_2i_4\cdot\cdot\cdot}^{j_0j_1\cdot\cdot\cdot}$'s,
$t_{i_2i_4\cdot\cdot\cdot}^{j_0j_1\cdot\cdot\cdot}$'s and
$u_{i_2i_4\cdot\cdot\cdot}^{j_0j_1\cdot\cdot\cdot}$'s refer to
arbitrary constant parameters. The positive integers
$i_2,i_4,i_6,\cdot\cdot\cdot$ run over the numbers 2 and 3,
the integer $j_0\geq1$, and the non-negative integers $j_0,j_1,j_2,\cdot\cdot\cdot$ are
constrained by $j_0+j_1+j_2+\cdot\cdot\cdot=i$. For convenience,
we have used the notations
\be
\mathrm{P}_1=\mathrm{P}(1,0,0) \, , \quad
\mathrm{P}_2=\mathrm{P}(0,1,0) \, , \quad
\mathrm{P}_3=\mathrm{P}(0,0,1)
\, , \label{P123}
\ee
while $\mathrm{P}_1^0=\mathrm{P}_2^0=\mathrm{P}_3^0$ denote identity operators.
By the way, it should be pointed out that the two operators $\mathrm{P}_2$ and
$\mathrm{P}_3$ have to be separated by $\mathrm{P}_1$ attributed to the fact that
$\mathrm{P}_2\mathrm{P}_3=0=\mathrm{P}_3\mathrm{P}_2$.
Apparently, the current $\textbf{J}_{HV}^{(2N)}$ includes $\textbf{J}_{(1)}$
in work \cite{Ppformin} as a special case. When $N=1$,
$\bar{\textbf{V}}_{i_2i_4\cdot\cdot\cdot}^{j_1j_2\cdot\cdot\cdot}=\textbf{V}$
and it accordingly becomes $\textbf{J}_{HV}^{(2)}=\textbf{J}_{V}$. In
contrast with the ordinary Komar current $\textbf{J}_{K}$, the co-closed
1-form $\tilde{\textbf{J}}_{HV}^{(2i)}$ is able to be viewed as the Noether
current corresponding to the diffeomorphism symmetry generated by the vector
field $\tilde{V}^\mu_{(i)}$, testifying the one-to-one correspondence between the
vector field and the conserved current  in another manner. Apart from this,
it may be interpreted as the higher-order derivative correction to
$\textbf{J}_{K}$. Therefore, $\tilde{\textbf{J}}_{HV}^{(2i)}$'s or
$\textbf{J}_{HV}^{(2N)}$ could have the applicability in computing the
conserved charges of higher-derivative extended gravity theories, such as
the well-known Lovelock gravity, and the so-called (generalized)
quasi-topological gravities.

As an example of Eq. (\ref{HCforV}), we specialize to the situation
where $N=2$. The 1-form $\textbf{J}_{HV}^{(2N)}$ accordingly turn into
\bea
\textbf{J}_{HV}^{(4)}&=&\textbf{J}_{V}
+k_{11}\mathrm{P}_1^2\textbf{V}
+k_{22}\mathrm{P}_2^2\textbf{V}
+k_{33}\mathrm{P}_3^2\textbf{V}\nn \\
&&+k_{12}\mathrm{P}_1\mathrm{P}_2\textbf{V}
+k_{13}\mathrm{P}_1\mathrm{P}_3\textbf{V} \nn \\
&&+k_{21}\mathrm{P}_2\mathrm{P}_1\textbf{V}
+k_{31}\mathrm{P}_3\mathrm{P}_1\textbf{V}
\, . \label{JHVk2}
\eea
Supposed that $\textbf{V}$ is the Killing vector $\boldsymbol{\xi}$, the
commutation relation given by Eq. (\ref{CRBKV}) enables us to write down
\bea
\textbf{J}_{HV}^{(4)}(\boldsymbol{\xi})&=&
(k_1+k_3)\hat{\delta}d\boldsymbol{\xi}
+\frac{1}{2}(k_{31}+2k_{33})\big(\hat{\delta}d\big)^2\boldsymbol{\xi} \nn \\
&&+(k_{11}+2k_{13})\Box^2\boldsymbol{\xi}
\, , \label{JHV2xi}
\eea
while $\mathbf{\mathcal{J}}_{HV}^{(4)}(\boldsymbol{\xi})
=\mathbf{\mathcal{J}}_{HV}^{(4)}(\textbf{V}\rightarrow\boldsymbol{\xi})$
can be expressed as \cite{PenZL}
\be
\mathbf{\mathcal{J}}_{HV}^{(4)}(\boldsymbol{\xi})
=\textbf{J}_{HV}^{(4)}(\boldsymbol{\xi})
+\alpha_{28}\hat{\delta}\Box d\boldsymbol{\xi}
\, . \label{CalJ2xi}
\ee
Apparently, the 1-forms $\textbf{J}_{HV}^{(4)}(\boldsymbol{\xi})$ and
$\mathbf{\mathcal{J}}_{HV}^{(4)}(\boldsymbol{\xi})$ are conserved
under the condition that
$(k_{11}+2k_{13})\hat{\delta}\Box^2\boldsymbol{\xi}=0$. Within \cite{PenZL},
it has been demonstrated that the surface integral with respect to
the potential derived from
$\mathbf{\mathcal{J}}_{HV}^{(4)}(\boldsymbol{\xi})$
can be used to define the mass and angular momentum of Einstein gravity
in asymptotically anti-de Sitter (AdS) spacetimes. If adding
a co-closed 1-form $\textbf{J}_{R}^{(4)}(\boldsymbol{\xi})$ deduced
from Eq. (\ref{GenJK}) to
$\mathbf{\mathcal{J}}_{HV}^{(4)}(\boldsymbol{\xi})$,
we are able to get a more general identically conserved current depending
at most upon the fourth-order derivatives of the Killing vector, that is,
\be
\hat{\textbf{J}}_{HV}^{(4)}(\boldsymbol{\xi})=
k_1\hat{\delta}d\boldsymbol{\xi}
+k_2\big(\hat{\delta}d\big)^2\boldsymbol{\xi}
+k_3\hat{\delta}\Box d\boldsymbol{\xi}
+\hat{\delta}\textbf{X}_{(2)}
+\textbf{J}_{R}^{(4)}(\boldsymbol{\xi})
\, . \label{JHVxiR}
\ee
In Eq. (\ref{JHVxiR}), the 2-form $\textbf{X}_{(2)}$ is defined by
\be
X_{(2)}^{\mu\nu}=2\Big(\beta_1 R^{\mu\nu}_{\rho\sigma}
+\beta_2 R^{[\mu}_{[\rho} \delta^{\nu]}_{\sigma]}
+\beta_3 R \delta^{[\mu}_{[\rho} \delta^{\nu]}_{\sigma]}\Big)
\nabla^\rho\xi^\sigma
\, , \label{X2def}
\ee
where $\beta_i$'s are arbitrary constant parameters. By utilizing
Eqs. (\ref{PoppForm}) and (\ref{DefOmega}), $\textbf{X}_{(2)}$ is rewritten as
\bea
\textbf{X}_{(2)}&=&
\beta_1d\hat{\delta}d\boldsymbol{\xi}
-\beta_1\Box d\boldsymbol{\xi}
+\beta_3 R d\boldsymbol{\xi} \nn \\
&&-(2\beta_1+\beta_2)R^\rho_{\mu}\nabla_{\nu}\xi_\rho
dx^\mu\wedge dx^\nu
\, . \label{X2def2}
\eea
Particularly, for the Einstein manifold with $R_{\mu\nu}=\lambda g_{\mu\nu}$,
we have $\textbf{X}_{(2)}=\beta_1d\hat{\delta}d\boldsymbol{\xi}
-\beta_1\Box d\boldsymbol{\xi}
+\lambda(2\beta_1+\beta_2+n\beta_3)d\boldsymbol{\xi}$.
The identically conserved
$\textbf{J}_{R}^{(4)}(\boldsymbol{\xi})$ is defined by
\bea
\textbf{J}_{R}^{(4)}(\boldsymbol{\xi})&=&
\big(a_0+a_1R+a_2\Box R+a_3R^2+a_4R_{\mu\nu}R^{\mu\nu}\nn \\
&&+a_5R_{\mu\nu\rho\sigma}R^{\mu\nu\rho\sigma}\big)\boldsymbol{\xi}
\, . \label{JRiem4th}
\eea
Here the coefficient of $\boldsymbol{\xi}$ in
$\textbf{J}_{R}^{(4)}(\boldsymbol{\xi})$ is a special case of $\lambda$ given
by Eq. (\ref{Sollam}), and it can be understood as the Lagrangian density of
the gravity theories with fourth-order derivatives. Thus the conserved current
(\ref{JHVxiR}) may be applicable in the definition of the conserved charges
for the no more than fourth-order derivative gravity theories, such as
the Einstein gravity, the Einstein-Weyl gravity and the Einstein-Gauss-Bonnet
gravity.

More generally, according to Eqs. (\ref{defW}) and (\ref{tiJVdef}), a conserved
current associated with no more than $(2N)$-th order derivatives of the Killing
vector $\boldsymbol{\xi}$ can be presented by
\be
\hat{\textbf{J}}_{HV}^{(2N)}(\boldsymbol{\xi})=
\sum_{i=1}^{N}\tilde{\textbf{J}}_{HV}^{(2i)}(\textbf{V}\rightarrow\boldsymbol{\xi})
+\boldsymbol{\xi}\sum_j a_j F_j
\, . \label{JHxi2k}
\ee
Here the scalars $F_i$'s are demanded to be dependent at most of $(2N)$-th
order derivatives of the metric tensor.

Furthermore, if the vector $V^\mu$ can not ensure the coderivatives of
the 1-forms $\textbf{J}_{V}$, $\tilde{\textbf{J}}_{V}$,
$\check{\textbf{J}}_{V}$, $\textbf{J}_{HV}^{(2N)}$ and
$\mathbf{\mathcal{J}}_{HV}^{(2N)}$
disappear, like in \cite{JuFeng}, we can introduce a scalar field
$\psi(x)$ to cancel out the non-vanishing divergence. For example,
when $\hat{\delta}\sum_{i=1}^N\bar{\textbf{J}}_{HV}^{(2i)}\neq0$,  we are able to modify
$\textbf{J}_{HV}^{(2N)}$ as the conserved current $\textbf{J}_{GHV}^{(2N)}$,
presented by
\be
\textbf{J}_{GHV}^{(2N)}=\textbf{J}_{HV}^{(2N)}+d\psi
\, . \label{HighCforCV}
\ee
Taking a divergence, we see that the scalar $\psi$ has to fulfill the wave
equation
\be
\Box\psi +\hat{\delta}\sum_{i=1}^N\bar{\textbf{J}}_{HV}^{(2i)}=0
\, . \label{DivJHV}
\ee
More concretely, by letting
$\psi=(\chi k_1+2k_2)\hat{\delta}\boldsymbol{\kappa}$,
the 1-form $\textbf{J}_{AKV}$ given by Eq. (\ref{CurraKV}) can be put into
the divergence-free form $(\textbf{J}_{AKV}-d\psi)$.

\section{Conclusions and remarks}\label{five}
By virtue of the quadratic differential operators $\Box$, $d\hat{\delta}$ and
$\hat{\delta}d$, together with their linear combination $\mathrm{P}(k_1,k_2,k_3)$
given by (\ref{OpefBhdd}), we have generalized
the usual Komar current to the ones for Killing vectors or generic vector fields.
Starting with the Komar conserved current (\ref{Komcuuf2}) expressed by the above operations,
we have derived the most general conserved current (\ref{ckJK}) under the condition
that the current without the linearity in the Killing vector field relies at
most upon the second-order derivatives of this vector, as well as its
generalization (\ref{GenJK}). Subsequently, the current (\ref{Komcuuf2}) is
extended to the one $\textbf{J}_{V}$ (\ref{CurrofGV}) for a generic
vector field constrained by Eq. (\ref{divofCofGV}). As some examples to
$\textbf{J}_{V}$, we have derived the conserved currents associated with a
couple of interesting specific vector fields, such as the almost-
and conformal Killing vectors, together with the affine collineation vector.
Remarkably, in terms of the combined operator $\mathrm{P}$, all the equations
involved in second-order derivatives of these vectors can be written as the
unified formulation (\ref{SpecV}). Furthermore, the operator $\mathrm{P}$
assists us to conveniently generalize $\textbf{J}_{V}$ to the conserved currents
$\mathbf{\mathcal{J}}_{HV}^{(2N)}$ and  $\textbf{J}_{HV}^{(2N)}$ given by
Eqs. (\ref{CalJ2Ndef}) and (\ref{HCforV}) respectively, which contain higher-order
derivatives of a general vector field and might be used to defined the conserved
charges of higher-derivative gravity theories. In order to understand the general
conserved currents $\mathbf{\mathcal{J}}_{HV}^{(2N)}$ and  $\textbf{J}_{HV}^{(2N)}$,
we propose that an effective approach is to compare it with the Noether
current obtained through the standard Noether method.

Apparently, the conserved currents in the present work are constructed
from the mathematical point of view. However, they have physical significance.
Among them, one or more may be expected to possess some nice properties beyond
the usual Komar current. For instance, although the conventional Komar
current breaks down in the definition for the mass of asymptotically AdS
black holes, in the work \cite{PenZL}, it has been shown that it corrected by
the fourth-order derivatives of the Killing vector can yield their physical
mass. Therefore, in order to understand these currents, some research desired
is to investigate the applications of the conserved currents in the
definition for the conserved
charges of various gravity theories, particularly of the higher-derivative
gravities. Noteworthily, when these currents are applied to compute the
charges of some specific gravity theory, one may encounter the problem
how to fix the coefficients in them for the sake of finding out the
appropriate conserved currents. Anyway, like in \cite{PenZL}, we think that such conserved currents
at least have to give rise to convergent charges that satisfy the first law of
thermodynamics or match those via other typical methods. We look forward to
the further investigations to answer all the mentioned problems.

\section*{Acknowledgments}

This work was supported by the Natural Science Foundation of China under Grant
Nos. 11865006 and 11505036. It was also partially supported by the Technology
Department of Guizhou province Fund under Grant Nos. [2018]5769 and [2016]1104.

\appendix
\section{Introduction to the operators $\Box$,
$\hat{\delta}d$ and $d\hat{\delta}$}\label{appendA}

Within the work \cite{Ppformin}, it has been demonstrated that both the two
fundamental operators $\hat{\delta}d$ and $d\hat{\delta}$, constructed
out of the combination of the Hodge star $\star$ and the exterior derivative
$d$, can be employed in the computations of the second-order derivatives
of an arbitrary $p-$form $\textbf{F}$ and maintain its form degree unchanged.
Here the Hodge star operation is defined through
$(\star \textbf{F})_{\mu_1\cdot\cdot\cdot\mu_q}=
(p!)^{-1}F^{\nu_1\cdot\cdot\cdot\nu_p}
\epsilon_{\nu_1\cdot\cdot\cdot\nu_p\mu_1\cdot\cdot\cdot\mu_q}$ under its action
on the $p-$form $\textbf{F}$, while the exterior differential of $\textbf{F}$ is
presented by
$(d\textbf{F})_{\mu_0\cdot\cdot\cdot\mu_p}
=(p+1)\nabla_{[\mu_0}F_{\mu_1\cdot\cdot\cdot\mu_p]}$.
In terms of the linear combination of $d\hat{\delta}$ and $\hat{\delta}d$, the
action of the ordinary generally covariant d'Alembertian operator
$\Box=g^{\mu\nu}\nabla_\mu\nabla_\nu$ on $\textbf{F}$, which leaves
the form degree unaltered as well, could be written as the so-called Weitzenb\"{o}ck
identity \cite{Ppformin}
\be
\Box\textbf{F}=d\hat{\delta}\textbf{F}+\hat{\delta}d\textbf{F}
-\mathbf{\Omega}(\textbf{F})
\, , \label{PoppForm}
\ee
where in component notation the $p$-forms $\Box\textbf{F}$,
$d\hat{\delta}\textbf{F}$, and $\hat{\delta}d\textbf{F}$ are given by
\bea
(\Box\textbf{F})_{\mu_1\cdot\cdot\cdot\mu_p}&=&
\Box F_{\mu_1\cdot\cdot\cdot\mu_p}
\, , \nn \\
(d\hat{\delta}\textbf{F})_{\mu_1\cdot\cdot\cdot\mu_p}
&=&p\nabla_{[\mu_1}\nabla^\nu F_{|\nu|\mu_2\cdot\cdot\cdot\mu_p]}
\, , \nn \\
(\hat{\delta}d\textbf{F})_{\mu_1\cdot\cdot\cdot\mu_p}
&=& \Box F_{\mu_1\cdot\cdot\cdot\mu_p}-p\nabla^\nu\nabla_{[\mu_1}
F_{|\nu|\mu_2\cdot\cdot\cdot\mu_p]}
\, , \label{ComBhdd}
\eea
respectively, while the component of $p$-form $\mathbf{\Omega}(\textbf{F})$
takes the form
\be
\big(\mathbf{\Omega}(\textbf{F})\big)_{\mu_1\cdot\cdot\cdot\mu_p}
=-pR^\sigma_{[\mu_1}
F_{\mid\sigma\mid\mu_2\cdot\cdot\cdot\mu_p]}+\frac{p(p-1)}{2}
R^{\rho\sigma}_{[\mu_1\mu_2}
F_{|\rho\sigma|\mu_3\cdot\cdot\cdot\mu_p]}
\, . \label{DefOmega}
\ee
Here $R_{\rho\sigma\mu\nu}$ denotes the standard Riemann-Christoffel tensor
of the spacetime metric, defined through
$2\nabla_{[\mu}\nabla_{\nu]}V_\rho=R_{\mu\nu\rho\sigma}V^\sigma$. In terms
of Eq. (\ref{ComBhdd}), we obtain
$(\hat{\delta}d)(d\hat{\delta})=(d\hat{\delta})(\hat{\delta}d)=0$,
as well as the commutation relations
$\big[d\hat{\delta},\Box\big]\textbf{F}=\mathbf{\Omega}(d\hat{\delta}\textbf{F})
-d\hat{\delta}\mathbf{\Omega}(\textbf{F})$ and
$\big[\hat{\delta}d,\Box\big]\textbf{F}=\mathbf{\Omega}(\hat{\delta}d\textbf{F})
-\hat{\delta}d\mathbf{\Omega}(\textbf{F})$. Particularly, in the case where
$\textbf{F}$ becomes the 1-form $\textbf{V}$, the former is concretely written
as
\be
\big[d\hat{\delta},\Box\big]\textbf{V}=
\frac{1}{2}\big(\nabla_\mu(V^\nu\nabla_\nu R)
+2\nabla_\mu(R_{\rho\sigma}\nabla^\rho V^\sigma)
-2R_{\mu\nu}\nabla^\nu\nabla^\sigma V_\sigma
\big)dx^\mu
\, , \label{CRBV1}
\ee
and the latter turns into
\bea
\big[\hat{\delta}d,\Box\big]\textbf{V}&=&
\big[(\nabla^\rho\nabla^\sigma R_{\rho\mu\sigma\nu}
-R^{\rho\sigma}R_{\rho\mu\sigma\nu}+R^\rho_\mu R_{\rho\nu})V^\nu \nn \\
&&+(\nabla^\nu R_{\nu\sigma\rho\mu}
+\nabla^\nu R_{\nu\mu\rho\sigma}
+\nabla_\sigma G_{\rho\mu})\nabla^\rho V^\sigma \nn \\
&&-R^{\sigma\nu}\nabla_\mu\nabla_\sigma V_\nu
+R_{\mu\nu}\nabla^\nu\nabla^\sigma V_\sigma
\big]dx^\mu
\, . \label{CRBV2}
\eea
When $\textbf{V}$ is the Killing vector $\boldsymbol{\xi}$,
Eqs. (\ref{CRBV1}) and (\ref{CRBV2}) further become respectively
\bea
d\hat{\delta}\Box\boldsymbol{\xi}&=&
\Box d\hat{\delta}\boldsymbol{\xi}=0 \, , \nn \\
\hat{\delta}d\Box\boldsymbol{\xi}&=&\Box\hat{\delta}d\boldsymbol{\xi}
-\Box^2\boldsymbol{\xi}+\mathbf{\Omega}(\Box\boldsymbol{\xi})
\, . \label{CRBKV}
\eea

Eq. (\ref{ComBhdd}) indicates
that $\Box$, $d\hat{\delta}$ and $\hat{\delta}d$ denote all the three elementary
second-order covariant derivative operators that can preserve the form degree
of an arbitrary $p$-form. To see this clearly, letting the covariant derivative operation
$\nabla_\mu$ act on a $p$-rank antisymmetric tensor $F_{\mu_1\cdot\cdot\cdot\mu_p}$
twice in succession, one finds that all the resultant $p$-rank antisymmetric
tensors have to take the three types of forms:
\bea
&&\nabla^\nu\nabla_\nu F_{\mu_1\cdot\cdot\cdot\mu_p} \, , \nn \\
&&\nabla_{[\mu_1}\nabla^\nu
F_{\mu_2\cdot\cdot\cdot\mid\nu\mid\cdot\cdot\cdot\mu_p]} \, , \nn \\
&&\nabla^\nu\nabla_{[\mu_1}
F_{\mu_2\cdot\cdot\cdot\mid\nu\mid\cdot\cdot\cdot\mu_p]}
\, . \label{Th2npF}
\eea
In the above equation, the first and second expressions are covered by the components
of $\Box\textbf{F}$ and $d\hat{\delta}\textbf{F}$ respectively, while the third one
in the notation of differential forms is just the linear combination of the $p$-forms $\Box\textbf{F}$ and
$\hat{\delta}d\textbf{F}$. Consequently, as what has been shown in
\cite{Ppformin}, instead of the operators
$\{\nabla^\nu\nabla_\nu,\nabla_\mu\nabla^\nu,\nabla^\nu\nabla_\mu\}$,
the three ones $\{\Box,d\hat{\delta},\hat{\delta}d\}$ can be always applied
to act on antisymmetric tensors because of their own advantages in differential
forms. Moreover, Eq. (\ref{Th2npF}) obviously shows that all the three operations
$\Box$, $d\hat{\delta}$ and $\hat{\delta}d$ are the fundamental ones of the
lowest differential order, which leave the form degree unchanged.

Additionally, for an arbitrary vector $V^\mu$, Eq. (\ref{PoppForm}) yields
the commutation relation between $\hat{\delta}$ and $\Box$, that is,
\be
\big[\hat{\delta},\Box\big]\textbf{V}=\frac{1}{2}V^\mu\nabla_\mu R
+R_{\mu\nu}\nabla^{(\mu} V^{\nu)}
\, . \label{CoBoCRV}
\ee
Apart from this, the more general commutation relation
$\big[\hat{\delta},\Box^m\big]\textbf{V}$ has been presented
in the work \cite{Ppformin}.

\section{The expression for $\textbf{J}_{V2}$}\label{appendB}

We assume that the 1-form $\textbf{J}_{V2}$ only depends on the terms
proportional to the second-order derivatives of the vector $V^\mu$, and
the number of $V^\mu$ in each term is permitted to be more than one. Hence,
the components of the 1-form $\textbf{J}_{V2}$ take the most general
form
\bea
(\textbf{J}_{V2})_\mu&=&
\lambda_{01}\nabla^\nu \Phi_{\mu\nu}
+\lambda_{02}\nabla_\mu\Phi
+\lambda_{03}\nabla^\nu \Psi_{\mu\nu}
+V^\nu\big(\lambda_{11}\Phi\Phi_{\mu\nu}
+\lambda_{12} \Phi\Psi_{\mu\nu}
 \nn \\
&&+\lambda_{13}\Phi_{\mu}^{\rho}\Phi_{\rho\nu}
+\lambda_{14}\Phi_{\mu}^{\rho}\Psi_{\rho\nu}
+\lambda_{15}\Phi_{\nu}^{\rho}\Psi_{\rho\mu}
+\lambda_{16}\Psi_{~\nu}^{\rho}\Psi_{\rho\mu}\big) \nn \\
&&+V^\rho V^\sigma
\big(\lambda_{21}\nabla_\mu\Phi_{\rho\sigma}
+\lambda_{22}\nabla_\sigma\Phi_{\rho\mu}
+\lambda_{23}\nabla_\rho\Psi_{\sigma\mu}\big)
\nn \\
&&+V^\nu V^\rho V^\sigma \big(\lambda_{31}\Phi_{\mu\nu}\Phi_{\rho\sigma}
+\lambda_{32}\Psi_{\mu\nu}\Phi_{\rho\sigma}\big)
+UV_\mu
\, , \label{JV2form}
\eea
with the scalar $U$ defined by
\bea
U&=&u_{01}\Phi^2
+u_{02}\Phi_{\rho\sigma}\Phi^{\rho\sigma}
+u_{03}\Psi_{\rho\sigma}\Psi^{\rho\sigma}
+V^\rho\big(u_{11}\nabla_\rho\Phi
+u_{12}\nabla^\sigma\Phi_{\rho\sigma}
+u_{13}\nabla^\sigma\Psi_{\rho\sigma}\big)
\nn \\
&&+V^\rho V^\sigma \big(u_{21}\Phi\Phi_{\rho\sigma}
+u_{22}\Phi_{\rho}^{\nu}\Phi_{\nu\sigma}
+u_{23}\Phi_{\rho}^{\nu}\Psi_{\nu\sigma}
+u_{24}\Psi_{~\rho}^{\nu}\Psi_{\nu\sigma}\big) \nn \\
&&+u_{31}V^\nu V^\rho V^\sigma \nabla_{\nu}\Phi_{\rho\sigma}
+u_{41}\big(\Phi_{\rho\sigma}V^\rho V^\sigma\big)^2
+u_{00}R
\, . \label{DefofU}
\eea
In Eqs. (\ref{JV2form}) and (\ref{DefofU}), the scalars
$\lambda_{ij}=\tilde{\lambda}_{ij}f_{ij}(V)$ and
$u_{ij}=\tilde{u}_{ij}h_{ij}(V)$. Here $\tilde{\lambda}_{ij}$'s and
$\tilde{u}_{ij}$'s stand for arbitrary constant parameters since it is
assumed that the vector field and the metric tensor, together with their
derivatives, are the only variables of the conserved currents throughout
the present work.

\section{Some properties of Killing vectors}\label{appendC}

In this appendix, we shall provide some important properties of the Killing
vector $\xi^\mu$, which are tightly relevant to our calculations.

In particular, when the differential form $\textbf{F}$ in Eqs. (\ref{PoppForm})
and (\ref{ComBhdd}) is the 1-form Killing vector field $\boldsymbol{\xi}$, according to both the equations,
one obtains the well-known relationship
$\Box\boldsymbol{\xi}=\mathbf{\Omega}(\boldsymbol{\xi})$, or
\be
\Box\xi_\mu+R_{\mu\nu}\xi^\nu =0
\, . \label{KVRrelas}
\ee
Additionally, as a special case where the 1-form $\textbf{V}$ in Eq. (\ref{CoBoCRV}) is just
the Killing vector field $\boldsymbol{\xi}$, in terms of the identity
$2\Box\boldsymbol{\xi}=\hat{\delta}d\boldsymbol{\xi}$ from Eq. (\ref{DivPh}),
giving rise to $\hat{\delta}\Box\boldsymbol{\xi}=0$, together with the
divergenceless equation $\hat{\delta}\boldsymbol{\xi}=0$,
Eq. (\ref{CoBoCRV}) leads to
\be
\xi^\mu\nabla_\mu R=0 \, , \quad
\nabla^\mu\big(\Box^2\xi_\mu+R_{\mu}^{\nu}R^{\sigma}_{\nu}\xi_\sigma\big)=0
\, , \label{xiRcont}
\ee
The first equation in Eq. (\ref{xiRcont}) can be written as
$\mathcal{L}_\xi R=0$. Here $\mathcal{L}_\xi$ denotes the
Lie derivative along the Killing vector $\xi^\mu$.

Letting the covariant derivative act on the Killing vector $\xi^\mu$ twice, one obtains
the following equation
\be
\nabla_\mu\nabla_\nu \xi_\rho=R_{\rho\nu\mu\sigma}\xi^\sigma
\, . \label{Cd2KV}
\ee
This is attributed to the fact that $\nabla_\mu\xi_\nu=-\nabla_\nu\xi_\mu$
and $d^2\boldsymbol{\xi}=0$ yields $\nabla_{[\mu}\nabla_\nu \xi_{\rho]}=0$.
Obviously, Eq. (\ref{KVRrelas}) can also arise from the contraction btween
the $\mu$ and $\nu$ indices in the above equation.
In light of Eq. (\ref{Cd2KV}), one further arrives at the identity
\be
\xi^\rho \big(\nabla_\mu\nabla_\nu \xi_\rho\big)\nabla^\mu\xi^\nu=0
\, , \label{xiCd2cdxi}
\ee
or equivalently,
\be
R_{\rho\mu\nu\sigma}\xi^\rho\xi^\sigma\nabla^\mu\xi^\nu =0
\, . \label{xiCd2cdxi2}
\ee
In addition, Eq. (\ref{xiCd2cdxi2}) together with (\ref{Cd2KV}) enables us to gain
\be
\xi^\rho\xi^\sigma\nabla^\mu\nabla_\rho\nabla_\sigma \xi_\mu=0
\, . \label{xi23Cd}
\ee
Expanding the above equation yields
\be
\xi_\rho \big(\nabla^\sigma\xi^\rho\big)\Box\xi_\sigma
=\xi^\rho\xi^\sigma\nabla_\rho\Box\xi_\sigma
\, , \label{xi23Cd2}
\ee
giving rise to
\be
\xi_\rho \nabla^\rho\big(\xi^\sigma\Box\xi_\sigma\big)=0
\, . \label{xi23Cd3}
\ee

It is worth noticing that the above equation can be deduced instead from
the one $\xi_\rho \nabla^\rho\big(\xi^\sigma\Box\xi_\sigma\big)
=-\mathcal{L}_\xi\big(R_{\mu\nu}\xi^\mu\xi^\nu\big)$ by virtue of
Eq. (\ref{KVRrelas}), while the latter vanishes since
$\mathcal{L}_\xi R_{\mu\nu}=0$ and $\mathcal{L}_\xi \xi^\mu=0$. Apart from
this, by making use of the commutation relation
$\mathcal{L}_\xi \nabla_{\mu}=\nabla_{\mu}\mathcal{L}_\xi$
between the Lie derivative along the Killing vector and the covariant derivative,
we are able to prove that
\be
\xi^\mu\nabla_\mu\big[\big(\nabla^\nu\xi^2\big)\big(\nabla_\nu\xi^2\big)\big]
= 2\big(\nabla^\nu\xi^2\big)\nabla_\nu\mathcal{L}_\xi\xi^2=0
\, . \label{Liedsqxi}
\ee
More generally, for the scalars $F_i$'s and $W_i$'s defined by Eq. (\ref{defW}), their Lie
derivatives along the Killing vector are read off as
\bea
\xi^\mu \nabla_\mu F_i&=&F_i\big(\mathcal{L}_\xi g\big)
+F_i\big(\mathcal{L}_\xi \mathcal{R}\big)
+F_i\big(\nabla\mathcal{L}_\xi \mathcal{R}\big)
+\cdot\cdot\cdot=0 \, , \nn \\
\xi^\mu \nabla_\mu W_i&=&W_i\big(\mathcal{L}_\xi g\big)
+W_i\big(\mathcal{L}_\xi\xi\big)
+W_i\big(\nabla\mathcal{L}_\xi\xi\big)
+\cdot\cdot\cdot=0
\, . \label{CdW}
\eea

Lastly, another  useful identity associated with the Killing vector field is
\be
\big(\nabla^\mu\xi^\nu\big)\big(\nabla^\rho\xi_\nu\big)
\big(\nabla_\rho\xi_\mu\big)=0
\, . \label{ThCdxi}
\ee

\section{Summarization of conserved currents}\label{appendD}

In this appendix, for convenience, we summarize our main results in
this paper, which are the conserved currents presented by Table
\ref{CCurrents}.

\begin{table}[H]
\caption{Conserved currents with respect to vector fields}
    \vspace{15pt}
    \centering
    \begin{tabular}{p{2.5cm}p{2.6cm}p{3.4cm}}
        \hline
        \hline
     Current  & In equation &Constraint     \\
        \hline
        $\textbf{J}_K$, $\tilde{\textbf{J}}_K$
        & (\ref{KomCurr}), (\ref{Komcuuf2})
        & None \\
        $\hat{\textbf{J}}_K$, $\check{\textbf{J}}_K$
        & (\ref{GenKoC}), (\ref{ckJK})
        & Eq. (\ref{divJbarK}) \\
        $\textbf{J}_{V}$, $\textbf{J}^Y_{V}$
        & (\ref{CurrofGV}), (\ref{CCuofYV})
        &Eqs. (\ref{divofCofGV}), (\ref{SpecV}) \\
        $\textbf{J}_{AKV}$
        &(\ref{CurraKV})
        &$(k_1\chi+2k_2)\hat{\delta}d\hat{\delta}\boldsymbol{\kappa}=0$ \\
        $\textbf{J}_{CKV}$
        &(\ref{CCforConV})
        &$(k_1+nk_2)\Box\phi =0$\\
        $\textbf{J}^{gr}_{\boldsymbol{\omega}}$
        &(\ref{JVomexi})
        &  None    \\
        $\tilde{\textbf{J}}_{V}$, $\check{\textbf{J}}_{V}$
        &(\ref{TildJV}), (\ref{JV12Coder})
        & $\hat{\delta}\tilde{\textbf{J}}_{V}$,
        $\hat{\delta}\check{\textbf{J}}_{V}=0$ \\
        $\mathbf{\mathcal{J}}_{HV}^{(2N)}$,
        $\bar{\mathbf{\mathcal{J}}}_{HV}^{(2i)}$
        &(\ref{CalJ2Ndef})
        & $\hat{\delta}\mathbf{\mathcal{J}}_{HV}^{(2N)}=0$\\
        $\textbf{J}_{HV}^{(2N)}$
        &(\ref{HCforV})
        &$\hat{\delta}\sum_{i=1}^{N}\bar{\textbf{J}}_{HV}^{(2i)}=0$ \\
        $\bar{\textbf{J}}_{HV}^{(2i)}$, $\tilde{\textbf{J}}_{HV}^{(2i)}$
        & (\ref{bJVdef}), (\ref{tiJVdef})
        &$\hat{\delta}\bar{\textbf{J}}_{HV}^{(2i)}=0$ \\
        $\mathbf{\mathcal{J}}_{HV}^{(4)}(\boldsymbol{\xi})$
        & (\ref{JHV2xi}), (\ref{CalJ2xi})
        & $(k_{11}+2k_{13})\hat{\delta}\Box^2\boldsymbol{\xi}=0$\\
        $\hat{\textbf{J}}_{HV}^{(4)}(\boldsymbol{\xi})$
        &  (\ref{JHVxiR})
        & None\\

      \hline
    \end{tabular}
    \label{CCurrents}
\end{table}


\begin{thebibliography}{100}

\bibitem{Komar}
A. Komar, Covariant conservation laws in general relativity,
Phys. Rev. \textbf{113} (1959), 934-936.


\bibitem{Katfac2}
J. Katz, A note on Komar's anomalous factor,
Class. Quantum Grav., \textbf{2} (1985), 423.

\bibitem{PeKf2}
A.N. Petrov and J. Katz,
Conserved currents, superpotentials and cosmological perturbations,
Proc. Roy. Soc. Lond. A \textbf{458} (2002), 319-337.

\bibitem{JuSi}
B. Julia and S. Silva,
Currents and superpotentials in classical gauge theories: 2. Global aspects and
the example of affine gravity,
Class. Quantum Grav. \textbf{17} (2000), 4733-4744.

\bibitem{PetFtc}
A.N. Petrov,
Field-theoretical construction of currents and superpotentials in
Lovelock gravity,
arXiv:1903.05500 [gr-qc].

\bibitem{JuFeng}
J.C. Feng, 	
Some globally conserved currents from generalized Killing vectors
and scalar test fields,
Phys. Rev. D \textbf{98} (2018), 104035.

\bibitem{LBBKofl}	
D. Lynden-Bell and J. Bi\v{c}\'{a}k,
Komar fluxes of circularly polarized light beams and cylindrical metrics,
Phys. Rev. D \textbf{96} (2017), 104053.

\bibitem{CleGal}
G. Cl\'{e}ment and D. Gal'tsov,
On the Smarr formula for rotating dyonic black holes,
Phys. Lett. B \textbf{773} (2017), 290-294.

\bibitem{KasKinHDG}
D. Kastor,
Komar integrals in higher (and lower) derivative gravity,
Class. Quantum Grav. \textbf{25} (2008), 175007.

\bibitem{BaZygrg}
A Gauss type law for gravity with a cosmological constant
S.L. Bazanski and P. Zyla,
Gen. Rel. Grav. \textbf{22} (1990), 379-387.

\bibitem{BNSnut}
G. Bossard, H. Nicolai and K.S. Stelle,
Gravitational multi-NUT solitons, Komar masses and charges,
Gen. Rel. Grav. \textbf{41} (2009), 1367-1379.

\bibitem{MielAgK}
E.W. Mielke,
Affine generalization of the Komar complex of general relativity,
Phys. Rev. D \textbf{63} (2001), 044018.

\bibitem{ACOTZ}
R. Aros, M. Contreras, R. Olea, R. Troncoso and J. Zanelli,
Conserved charges for gravity with locally AdS asymptotics,
Phys. Rev. Lett. \textbf{84} (2000), 1647-1650.

\bibitem{BFFV}
A. Borowiec, M. Ferraris, M. Francaviglia and I. Volovich,
Universality of Einstein equations for the Ricci squared Lagrangians,
Class. Quantum Grav. \textbf{15} (1998), 43-55.

\bibitem{GiaSa}
G. Giachetta and G. Sardanashvily,
Stress energy momentum of affine metric gravity. Generalized Komar
superpotential,
Class. Quantum Grav. \textbf{13} (1996), L67-L72.

\bibitem{MaKiAdS}
A. Magnon,
On Komar integrals in asymptotically anti-de Sitter space-times,
J. Math. Phys. \textbf{26} (1985), 3112-3117.

\bibitem{TamWin}
L.A. Tamburino, J.H. Winicour,
Gravitational fields in finite and conformal Bondi frames,
Phys. Rev. \textbf{150} (1966), 1039-1053.

\bibitem{Ppformin}
J.J. Peng,
Constructing p,n-forms from p-forms via the Hodge star operator
and the exterior derivative,
Commun. Theor. Phys. \textbf{72} (2020), 065402.

\bibitem{RodWai}
W.A. Rodrigues and S.A. Wainer,
Notes on conservation laws, equations of motion of matter, and particle
fields in Lorentzian and teleparallel de Sitter space-time structures,
Adv. Math. Phys. \textbf{2016} (2016), 5465263.

\bibitem{CHTau}	
C.H. Taubes,
Solution of the almost-Killing equation and conformal
almost-Killing equation in the Kerr spacetime,
J. Math. Phys. \textbf{19} (1978), 1515-1525.

\bibitem{BoCaPaL}
C. Bona, J. Carot, and C. Palenzuela-Luque,
Almost-stationary motions and gauge conditions in general relativity,
Phys. Rev. D \textbf{72} (2005), 124010.

\bibitem{RuPaBo}
M. Ruiz, C. Palenzuela, and C. Bona,
Almost-Killing conserved currents: a general mass function,
Phys. Rev. D \textbf{89} (2014), 025011.

\bibitem{KaRTinJ}
D. Kastor, S. Ray and J. Traschen,
Enthalpy and the mechanics of AdS black holes,
Class. Quantum Grav. \textbf{26} (2009), 195011.

\bibitem{IyerWald}
V. Iyer and R.M. Wald,
Some properties of the Noether charge and a proposal for dynamical black hole entropy,
Phys. Rev. D \textbf{50} (1994), 846.

\bibitem{PengN}
J.J. Peng,
Conserved charge of gravity theory with p-form gauge fields and its property
under Kaluza-Klein reduction,
Phys. Rev. D, \textbf{95} (2017), 104022.

\bibitem{sKVK}
A. Komar,
Asymptotic covariant conservation laws for gravitational radiation,
Phys. Rev. \textbf{127} (1962), 1411-1418.

\bibitem{MaASSG}
R.A. Matzner,
Almost symmetric spaces and gravitational radiation,
J. Math. Phys. \textbf{9} (1968), 1657-1668.

\bibitem{AppKB}
C. Beetle, Approximate Killing fields as an eigenvalue problem,
arXiv:0808.1745 [gr-qc].

\bibitem{PenZL}
J.J. Peng, C.L. Zou and H.F. Liu,
A Komar-like integral for mass and angular momentum
of asymptotically AdS black holes in Einstein gravity,
Phys. Scr. \textbf{96} (2021), 125207.

\end{thebibliography}
\end{document}